\documentclass[a4paper,useAMS,aasms,astrobib,usenatbib]{mn2e}
\usepackage{epsfig,amssymb,graphicx,rotating,color,etoolbox}
\usepackage[colorlinks,linkcolor=blue,citecolor=blue,urlcolor=blue]{hyperref}        
\usepackage[all]{hypcap}
\makeatletter
\patchcmd{\NAT@citex}
  {\@citea\NAT@hyper@{%
     \NAT@nmfmt{\NAT@nm}%
     \hyper@natlinkbreak{\NAT@aysep\NAT@spacechar}{\@citeb\@extra@b@citeb}%
     \NAT@date}}
  {\@citea\NAT@nmfmt{\NAT@nm}%
   \NAT@aysep\NAT@spacechar\NAT@hyper@{\NAT@date}}{}{}
\patchcmd{\NAT@citex}
  {\@citea\NAT@hyper@{%
     \NAT@nmfmt{\NAT@nm}%
     \hyper@natlinkbreak{\NAT@spacechar\NAT@@open\if*#1*\else#1\NAT@spacechar\fi}%
       {\@citeb\@extra@b@citeb}%
     \NAT@date}}
  {\@citea\NAT@nmfmt{\NAT@nm}%
   \NAT@spacechar\NAT@@open\if*#1*\else#1\NAT@spacechar\fi\NAT@hyper@{\NAT@date}}
  {}{}
\makeatother
\usepackage{txfonts}
\usepackage[T1]{fontenc}
\usepackage{aecompl}

\title[Eccentric circumplanetary discs]{1D accretion discs around eccentric planets:\\observable near-infrared variability}
\author[A. C. Dunhill]{A. C. Dunhill\textcolor{blue}{\thanks{E-mail: \href{mailto:adunhill@astro.puc.cl}{adunhill@astro.puc.cl}}}\\
Instituto de Astrof\'isica, Pontificia Universidad Cat\'olica de Chile, Vicu\~na Mackenna 4860, 7820436 Macul, Santiago, Chile}
\begin{document}
\voffset=-0.25in

\date{Accepted 2014 December 24. Received 2014 December 21; in original form 2014 November 04}

\pagerange{\pageref{firstpage}--\pageref{lastpage}} \pubyear{2014}

\maketitle

\label{firstpage}

\begin{abstract}
I present the results of 1D models of circumplanetary discs around planets on eccentric orbits. I use a classical viscous heating model to calculate emission fluxes at the wavelengths targeted by the NIRCam instrument on \textit{JWST}, and compare the variability of this signal with the published NIRCam sensitivity specifications. This variability is theoretically detectable by \textit{JWST} for a sufficiently viscous disc ($\alpha \sim 10^{-2}$) around a sufficiently eccentric planet ($e \sim 0.1-0.2$) and if the circumplanetary disc accretes material from its parent disc at a rate $\dot{M} \gtrsim 10^{-7}\,\mathrm{M}_{\sun}$ yr$^{-1}$. I discuss the limitations of the models used, and the implications of the result for probing the effectiveness of disc interactions for growing a planet's orbital eccentricity.
\end{abstract}


\section{Introduction}\label{sec:intro}

Planets form in and from gaseous circumstellar discs. In the final stages of formation, giant planets must accrete large amounts of gas from this disc. In order to allow the gas to accrete onto the planet, a circumplanetary disc (CPD) forms which acts as a bottleneck for this accretion, extracting angular momentum from the infalling gas. It has been recently shown that these discs may be directly observable in the near future with \textit{ALMA} \citep{wolfdangelo05,isellaetal14}, and also in the near infrared \citep{zhu14}. They therefore present an excellent opportunity for testing theoretical predictions of how accretion disc physics operates at these scales, and by inference how this is dictated by conditions in the wider circumstellar disc in which the CPD resides.

There has been much work in recent years exploring how CPDs form and evolve. Hydrodynamical simulations have been particularly useful in exploring the effect of disc viscosity \citep{buetal13,szulagyietal14} and different equations of state \citep{ayliffebate09b,gresseletal13}, but 1D models have also been widely employed \citep{martinlubow11,keithwardle14,zhu14}.

There is broad agreement from these simulations about the radial extent of a CPD due to tidal truncation \citep[$R_{\mathrm{out}} \sim 0.4\,R_{\mathrm{Hill}}$;][]{ayliffebate09b,martinlubow11}, and the effect of realistic thermodynamic treatment on this \citep[reducing the truncation radius by a factor of a few;][]{ayliffebate09b,gresseletal13}. CPDs are also expected to have high aspect ratios, with $H/R \sim 0.3 - 0.6$ depending again upon thermodynamic treatment \citep{ayliffebate09b,gresseletal13}.

However, there is still much work to be done on characterising the dynamical evolution of CPDs. Aspects such as the temperature and viscosity, for example, are the subject of much debate \citep[e.g.][]{szulagyietal14,gresseletal13,keithwardle14} as we know even less about the conditions to expect in the vicinity of a forming protoplanet than we do about the conditions in the wider protoplanetary disc, which is little enough \citep[e.g.][]{armitage11}.

An exciting possibility is that the observability of CPDs can give insights into the effect of resonant interactions between planets and their parent discs. Locally-isothermal simulations have been able to show this process growing the eccentricity of a planet in certain cases \citep*[e.g.][]{papaloizouetal01,dangeloetal06}, but they have also been shown to damp the eccentricity in cases where growth does not occur \citep*{dunhilletal13}. Proper treatment of the disc thermodynamics shows that this binarity (either growth or damping) is real \citep*{tsang14,tsangetal14}, but it is unclear which side of this fence protoplanetary discs sit on.

An ideal way to break this degeneracy would be to observe an eccentric planet embedded in a protoplanetary disc. Any such planet is highly likely to have grown its eccentricity in this way, as otherwise its eccentricity would have been damped. Recently \citet{zhu14} has calculated SEDs for the emission from a CPD around a forming planet. This takes the form of an excess above the star and circumstellar disc SEDs, which have been well studied and characterised \citep[e.g.][]{kenyonhartmann87,chianggoldreich97,zhuetal07}. It is possible that if the planet was on an eccentric orbit, the CPD's contribution to the SED would oscillate. This may allow us to directly identify an eccentric giant planet still forming.

In this Letter I use a simple 1D model of a CPD, and modulate the accretion of gas onto the CPD in a manner consistent with how eccentric planets accrete. I then model the emitted flux of the resultant disc over time, using an assumption of emission from viscous heating in the CPD, and compare the level of variability with the promised sensitivity of the NIRCam instrument on \textit{JWST}.

\section{Numerical model}\label{sec:sims}

I adopt the 1D numerical model described by \citet{martinlubow11}, which I shall briefly describe here\footnote{Strictly I do not adopt the formula for $\Omega$ used by \citet{martinlubow11}, instead using Keplerian values.}. This method evolves the 1D viscous diffusion equation, modified to account for tidal torques and  mass accretion onto the disc:
\begin{equation}
\frac{\upartial\Sigma}{\upartial t} =\frac{1}{R}\frac{\upartial}{\upartial R} \left[3R^{1/2}\frac{\upartial}{\upartial R} \left(\nu\Sigma R^{1/2}\right) - 2\Sigma\Omega^{-1}\frac{dT_{\mathrm{gr}}}{dM}\right] + S(R)
\label{eq:1}
\end{equation}
where $R$ is the radial distance from the planet, $\Sigma$ is the surface density in the circumplanetary disc and $\Omega$ is the Keplerian orbital frequency. $\nu$ is the kinematic viscosity in the disc, and I assume a \citet{shakurasunyaev73} $\alpha$ viscosity such that $\nu = \alpha H^2 \Omega$, where the scale height $H$ is set from the aspect ratio $H/R = 0.3$ and the $\alpha$ parameter is an input to be varied between models. $dT_{\mathrm{gr}}/dM$ is the tidal force truncating the outer edge of the circumplanetary disc and $S(R)$ is a source function representing the accretion from the circumstellar disc onto the CPD. For these functions I adopt the same form used by \citet{martinlubow11}, i.e. that
\begin{equation}
\frac{dT_{\mathrm{gr}}}{dM} = -\left(\frac{R}{R_{\mathrm{Hill}}}\right)^{4} \left \{ 
\begin{array}{ll}
0.5 R^2_{\mathrm{Hill}} \Omega^2_{\mathrm{p}} & R \geq 0.4 R_{\mathrm{Hill}}\\
0& \mathrm{otherwise}.
\end{array}
\right.
\label{eq:2}
\end{equation}
for a planet of Hill radius $R_{\mathrm{Hill}}$, and
\begin{equation}
S(R) = \frac{\dot{M}_{\mathrm{inj}}}{2\pi R_{\mathrm{inj}}}\,\frac{f\left[\left(R - R_{\mathrm{inj}}\right) / w\right]}{2w},
\label{eq:3}
\end{equation}
where $\dot{M}_{\mathrm{inj}}$ and $R_{\mathrm{inj}} = 0.2R_{\mathrm{Hill}}$ are the rate and radius at which mass is injected, $f(x) = 1$ for $|x| < 1$ or $0$ otherwise, and the injection width $w = 0.0046\sqrt{R_{\mathrm{inj}} R_{\mathrm{Hill}}}$.

Instead of using a static value for the Hill radius $R_{\mathrm{Hill}} = a(M_{\mathrm{\star}}/3M_{\star})^{1/3}$ for semimajor axis $a$, stellar mass $M_{\star}$ and planet mass $M_{\mathrm{p}}$, I use $R_{\mathrm{Hill}} = R_{\mathrm{sep}}\,(M_{\mathrm{\star}}/3M_{\star})^{1/3}$
in Equations \ref{eq:2} and \ref{eq:3}, where $R_{\mathrm{sep}}$ is the instantaneous separation between the planet and star, which in the limit of low eccentricity is well approximated by a sinusoid. This allows a rough approximation to the effects of an eccentric orbit, as the changing potential will affect truncation radius of a real CPD.

I use a fixed grid equispaced in $R^{1/2}$ with $370$ cells and evolve the equations using an explicit scheme \citep*[e.g.][]{pringleetal86}. Initially, I evolve the system for one viscous time $t_{\nu} \simeq R^2/\nu$ at $R=R_{\mathrm{Hill}}$ to allow the disc to settle into a steady state, with a constant $\dot{M}_{\mathrm{inj}}$ and $R_{\mathrm{sep}}$. This represents a planet on a circular orbit, and gives CPD profiles matching those of Figure 3 from \citet{martinlubow11}.

\subsection{Eccentricity}\label{sims:eccentricity}

After this stage, I begin to steadily increase the planet's eccentricity from 0 to $e$, another input parameter of the model. As the eccentricity increases $R_{\mathrm{sep}}$ becomes sinusoidal. I also begin to vary $\dot{M}_{\mathrm{inj}}$ in a similar manner. To parameterise how the injection rate should vary with eccentricity I use 3D SPH simulations of a 5 $M_{\mathrm{Jup}}$ planet with eccentricities $e = 0$, $0.05$ and $0.1$. The disc initial conditions and SPH code are identical to those of \citet*{dunhilletal13}, but with these eccentricities. In these simulations, the disc has a \citep{shakurasunyaev73} $\alpha$-viscosity with $\alpha = 0.01$. Although the equation of state is locally isothermal (meaning that the simulations are scale-free) the planet has a nominal semi-major axis $a_{\mathrm{p}} = 1$ au, and the disc has aspect ratio $H/R = 0.05$ and surface density $\Sigma = 100$ g cm$^{-2}$ at $R = a_{\mathrm{p}}$.

\begin{figure}
\includegraphics[width=\columnwidth]{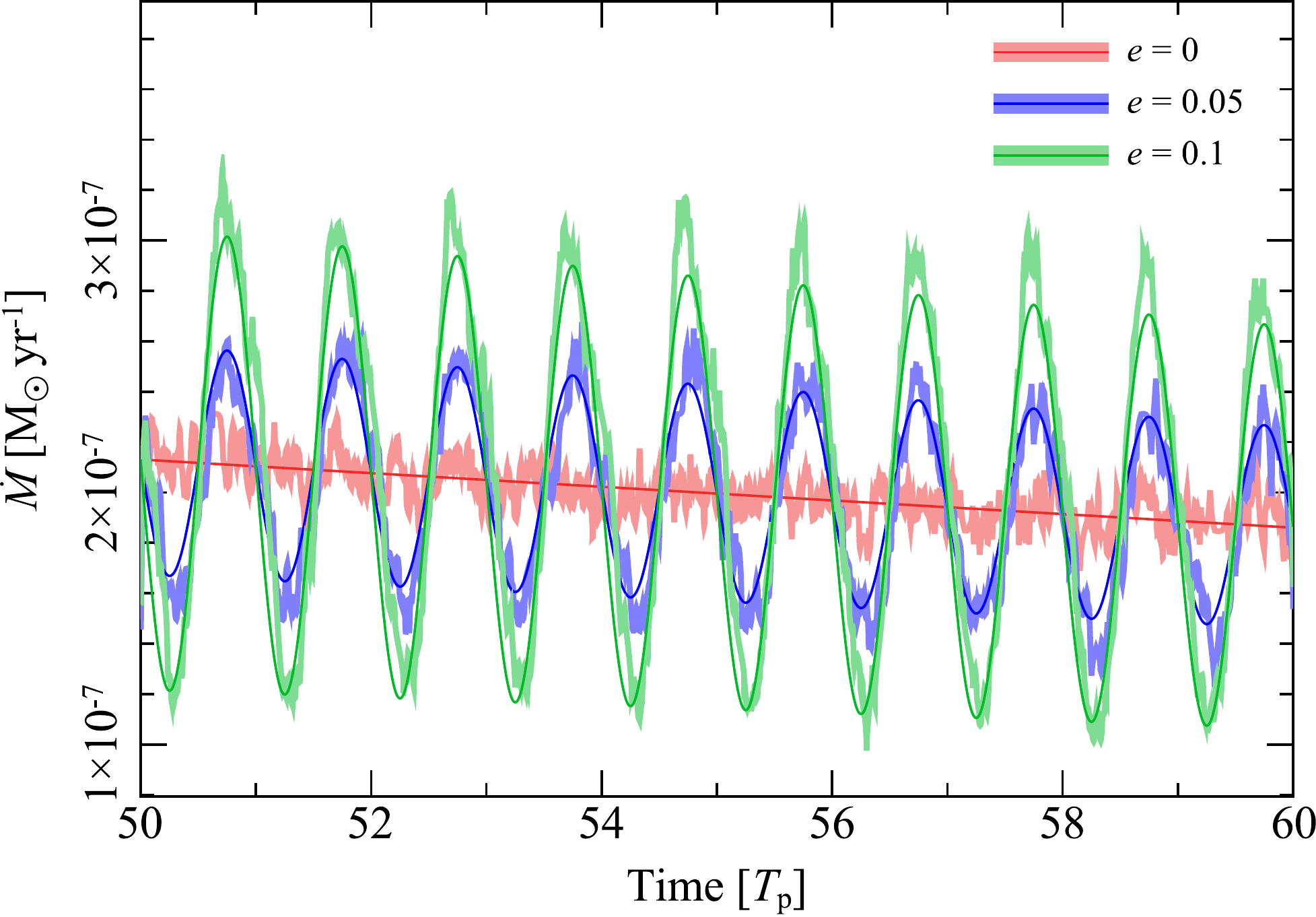}
\caption{Accretion rates onto $5 M_{\mathrm{Jup}}$ planets measured from high resolution SPH simulations for a eccentricities $e = 0$ (bold, pale red line), $e = 0.05$ (bold, pale blue line) and $e = 0.1$ (bold, pale green line). For the noneccentric orbit, the accretion rate is well represented by a straight (thin red) line, and for the eccentric planets the accretion rates are well fit by Equation \ref{eq:4} (thin blue and green lines for  $e = 0.05$ and  $e = 0.1$ respectively). As the disc is still relaxing at this point in the simulation, the accretion rate for the non-eccentric planet is decreasing.}
\label{fig:1}
\end{figure}

Although these simulations only resolve down to 0.4 $R_{\mathrm{Hill}}$, this is adequate to explore how the eccentricity affects the accretion rate onto the planet. Figure \ref{fig:1} shows these accretion rates after 50 planetary orbits, after the initial conditions have settled and the planet has opened a clean gap in the disc, using physical units corresponding to those noted above. Despite the noise in the SPH simulation, accretion rates on to the eccentric planets are well fit by
\begin{equation}
\dot{M}_{\mathrm{inj}}(e) = \dot{M}_{\mathrm{circ}} - \left[4.286\,e \dot{M}_{\mathrm{circ}}\,\mathrm{sin}\left(2\pi\,T/T_{\mathrm{p}}\right)\right]
\label{eq:4}
\end{equation}
where $\dot{M}_{\mathrm{circ}}$ is the accretion rate onto the non-eccentric planet. The accretion rate peaks when $R = a$ and the planet is half way between apocentre and pericentre (when $R_{\mathrm{sep}}$ is decreasing), and is at a minimum half an orbit later when between pericentre and apocentre (when $R_{\mathrm{sep}}$ is increasing). This is easily understood as the tidal streams delivering mass onto the planet are not in equilibrium for an eccentric planet, and they get ahead of the planet when the distance between the planet and star is decreasing, and the planet is able to catch up to the stream when the reverse is true, resulting in accretion minima and maxima respectively.

I therefore use Equation \ref{eq:4} to modulate $\dot{M}_{\mathrm{inj}}$ in the 1D model, with the appropriate $\pi/2$ offset between the orbital separations and $\dot{M}$. I steadily increase the eccentricity from 0 to $e$ over a viscous time in order to avoid unphysical transients caused by suddenly introducing the effect of eccentricity. I then keep $e$ constant for another viscous time before evaluating the disc over a final 50 orbital periods of the planet, by which time a steady state of constant variability has been reached. The values chosen for $\dot{M}_{\mathrm{circ}}$ in Equation \ref{eq:4} ($10^{-8}$ and $10^{-7}\,\mathrm{M}_{\sun}$ yr$^{-1}$) are based on values from SPH simulations by \citet{ayliffebate09a}, who found accretion rates in this range for Jupiter-mass planets, and on the MRI simulations by \citet{gresseletal13} who found very good agreement with these values. Although $\dot{M}$ onto the circular planet is decreasing in Figure \ref{fig:1}, this is for numerical reasons and in the 1D models $\dot{M}_{\mathrm{circ}}$ is constant.

It is unclear from simulations of embedded eccentric planets what the limiting value of $e_{\mathrm{max}}$ should be. \citet{papaloizouetal01} found growth of eccentricity due to resonant disc torques to values $e \sim 0.2$ for extremely massive bodies, but for planetary masses ($1 \leq M_{\mathrm{p}} \leq 10\,M_{\mathrm{Jup}}$) only up to $e \sim 0.05$, whereas \citet{dangeloetal06} found growth to $e \sim 0.1$.

The simplest limit on the eccentricity of a gap-opening planet is the width of the gap $w_{\mathrm{gap}}$: if the difference between apocentre and pericentre distances $R_{\mathrm{apo}} - R_{\mathrm{peri}} \gtrsim w_{\mathrm{gap}}$ then the interaction with the high density gas at the gap edge will effectively damp the eccentricity beyond some critical value \citep[analogous to the same eccentricity-limiting mechanism found for binaries by][]{roedigetal11}. \citet{cridaetal06} estimate that when a gap is opened, its half-width should be of order $2 R_{\mathrm{Hill}}$, corresponding to a critical eccentricity of $e \sim 0.2$ for a 5 $M_{\mathrm{Jup}}$ planet. I therefore test eccentricities up to this value in my models. The list of parameters which I vary between models and the values taken are listed in Table \ref{tab:1}.

\begin{table}
\begin{minipage}[t]{\columnwidth}\centering
\caption{Input parameters for the 1D disc models and their values.}\label{tab:1}
\begin{tabular}{lcc}
\hline
Parameter & Units & Values \\
\hline
$M_{\mathrm{p}}$ & $M_{\mathrm{Jup}}$ & $1$, $5$, $10$\\
$e$ & -- & $0.1$, $0.2$\\
$\alpha$ & -- & $10^{-3}$, $10^{-2}$\\
$\dot{M}_{\mathrm{circ}}$ & $\mathrm{M}_{\sun}$ yr$^{-1}$ & $10^{-8}$, $10^{-7}$\\
\hline
\end{tabular}
\end{minipage}
\end{table}

\section{Results}\label{sec:results}

This toy model of a CPD around an eccentric planet, although lacking in a number of ways (see Section \ref{discussion:limitations}), allows rudimentary estimates for how periodic modulation in the accretion onto the disc affect the its potential luminosity. To do this, I assume that energy dissipated by the disc viscosity is radiated away with 100 per cent efficiency as an accretion luminosity. The rate of dissipation in the disc per unit area at radius $R$ is given by
\begin{equation}
D(R) = \frac{1}{2}\nu\Sigma\left(R\frac{d\Omega}{dR}\right)^2.
\label{eq:5}
\end{equation}
This gives a disc surface temperature 
\begin{equation}
T_{\mathrm{s}} = \left(\frac{D(R)}{2\sigma}\right)^{1/4}
\label{eq:6}
\end{equation}
where $\sigma$ is the Stefan-Boltzmann constant \citep{pringle81}.

The flux $F_{\lambda}$ at wavelength $\lambda$ is then given by summing over disc annuli, each of which emits as a black body of temperature $T_{s}(R)$, so that
\begin{equation}
\lambda F_{\lambda} = \frac{1}{d^2} \int^{R_{\mathrm{out}}}_{R_{\mathrm{in}}} 2 \pi R \lambda B_{\lambda}\left(T_{\mathrm{s}}\right)\,dR
\label{eq:7}
\end{equation}
for a face-on disc at distance $d$ where $B_{\lambda}$ is the Planck function. Note that this neglects any contribution from direct mass accretion onto the planet (magnetospheric accretion), as \citet{zhu14} showed that this is likely negligible unless the planet's magnetic field is unrealistically strong (see Section \ref{discussion:limitations}).

\begin{figure}
\includegraphics[width=\columnwidth]{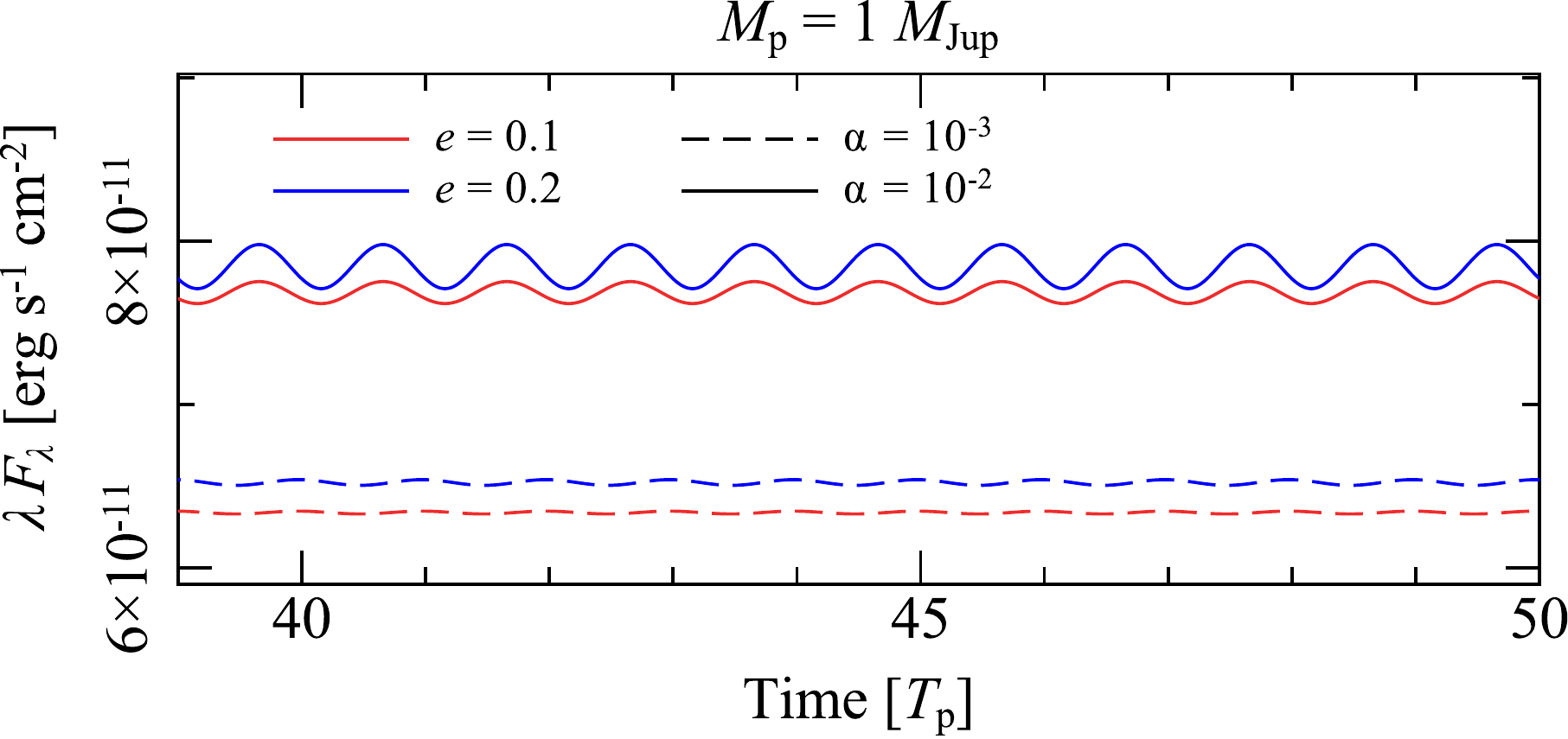}
\caption{Emitted flux $\lambda F_{\lambda}$ for a disc around a 1 $M_{\mathrm{Jup}}$ planet for $e = 0.1$ and $0.2$ (red and blue respectively) and \citeauthor{shakurasunyaev73} $\alpha = 10^{-3}$ and $10^{-2}$ (dashed and solid lines respectively) at $\lambda = 1.5\,\mu$m over the final eleven orbital periods of the planet $T_{\mathrm{p}}$. These models have $\dot{M}_{\mathrm{circ}} = 10^{-7}$. The period of the flux variability is the same as the planet's orbital period $T_{\mathrm{p}}$.}
\label{fig:2}
\end{figure}

As an example I plot the flux $\lambda F_{\lambda}$ from a CPD around a 1 $M_{\mathrm{Jup}}$ at an example wavelength of $1.5\,\mu$m as a function of time for eleven planetary orbits $T_{\mathrm{p}}$ for different eccentricities and disc viscosities and for $\dot{M}_{\mathrm{circ}} = 10^{-7}$ in Figure \ref{fig:2}. The variability produced by the eccentricity is the same as the planet's orbital period $T_{\mathrm{p}}$ independent of eccentricity, disc viscosity or wavelength. The viscosity produces large differences in the level of emission, as should be clear from Equations \ref{eq:5} to \ref{eq:7}, but the eccentricity also plays a large role in setting the amplitude of the modulation, $\Delta \lambda F_{\lambda}$.

in Figures \ref{fig:3} and \ref{fig:4} I plot the amplitude of this periodic variability in $\lambda F_{\lambda}$ for a subset of my models. The wavelengths chosen are those targeted by \textit{JWST}'s NIRCam instrument, for which the minimum instrument sensitivity specifications are also shown for comparison.  I choose the NIRCam wavelengths because they are in the spectral range where CPDs emit brightly \citep[$0.5-6\,\mu$m;][]{zhu14}\footnote{\textit{JWST} NIRCam photometric sensitivities taken from \href{http://www.stsci.edu/jwst/science/sensitivity/jwst-phot}{www.stsci.edu/jwst/{\\}science/sensitivity/jwst-phot}, and are the minimum design specifications.}. I take $d = 55$ pc, the distance of the TW Hya association, the nearest group of young potentially planet-forming discs to us.

These figures show that the flux variability induced by a planet's eccentricity is above the minimum sensitivity required for the NIRCam instrument, given a sufficiently viscous CPD, a high enough eccentricity and/or a high enough rate of accretion onto the CPD. It is unclear what the limiting values are for these parameters for a real protoplanet, but they are all within current uncertainties.

\section{Discussion}\label{sec:discussion}

\subsection{Limitations and omissions}\label{discussion:limitations}

This treatment  neglects non-axisymmetric effects, and the assumption of a Keplerian disc is not accurate, both due to the potential of the star and the thickness of the CPD. The 1D accretion disc diffusion equation (upon which Equation \ref{eq:1} is based) is only valid for thin discs where $H \ll R$. With an aspect ratio of $H/R = 0.3$, this is not the case here, and it must be noted that thick discs such as these have pressure gradients that strongly affect the rotation \citep{pringle81,lodato07}. I stick to the Keplerian assumption due to its simplicity, but note that it will be necessary to follow this up with a full 3D hydrodynamic treatment in future.

I also assume that the variation of $R_{\mathrm{sep}}$ over the course of the orbit of an eccentric planet is a perfect sinusoid. This is valid at low eccentricity, but becomes a poor approximation at $e = 0.2$. This may affect the shape of the light curves produced (for example that in Figure \ref{fig:2}) but should not affect the level of variability.

\begin{figure}
\includegraphics[width=\columnwidth]{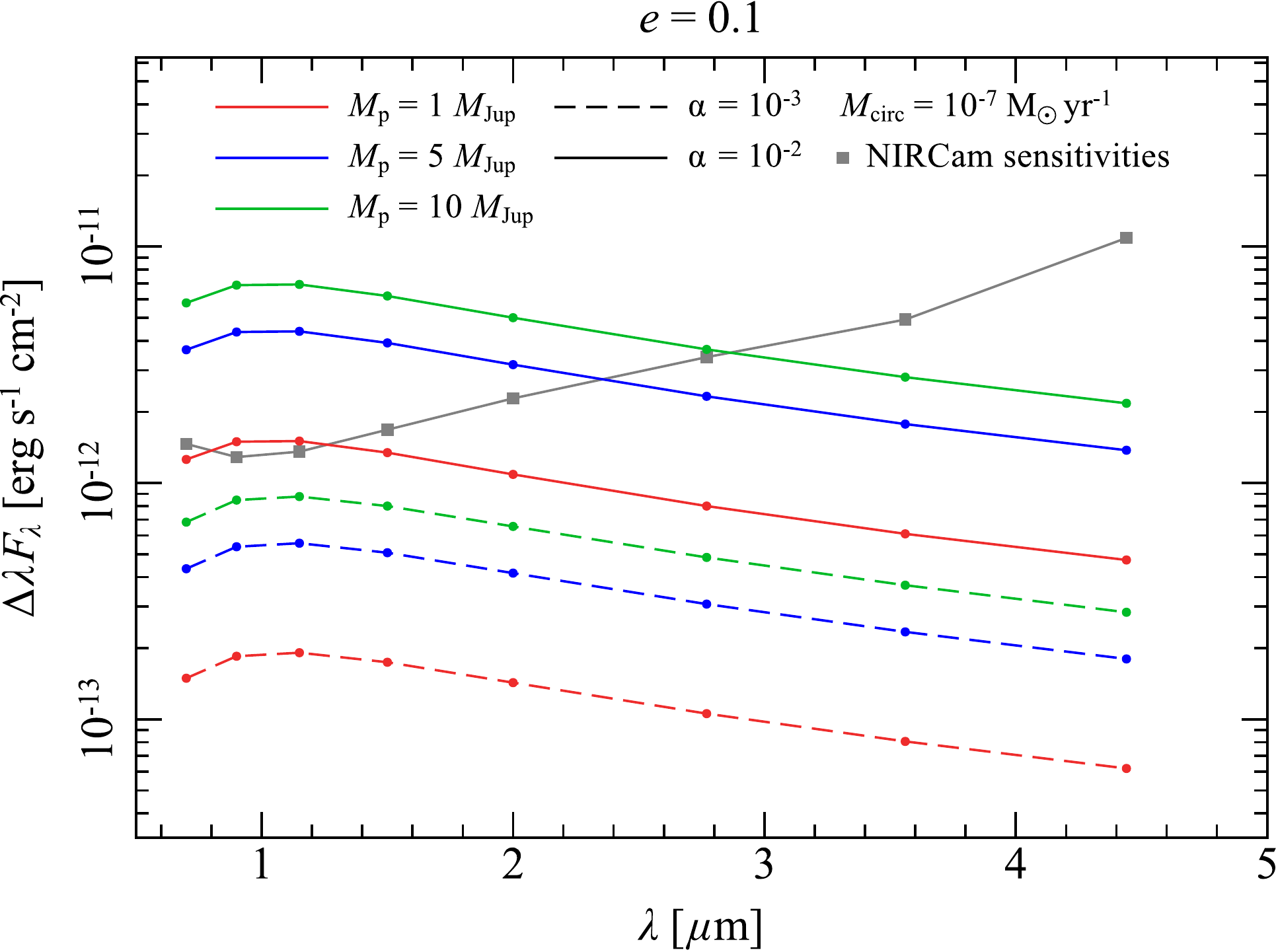}
\caption{Amplitude of the flux variation $\Delta \lambda F_{\lambda}$ against wavelength for different CPD models around planets with eccentricity $e = 0.1$, at a distance $d = 55$ pc. The models shown have a high accretion rate onto the disc, $\dot{M}_{\mathrm{circ}} = 10^{-7}\,\mathrm{M}_{\sun}$ (Equation \ref{eq:4}). Colour indicates planet mass ($M_{\mathrm{p}} = 1\,M_{\mathrm{Jup}}$, $5\,M_{\mathrm{Jup}}$ and $10\,M_{\mathrm{Jup}}$ are red, blue and green respectively), and dashed and solid lines indicates CPD viscosity ($\alpha = 10^{-3}$ and $\alpha = 10^{-2}$ respectively). Grey squares give the minimum \textit{JWST} NIRCam specification sensitivities.}
\label{fig:3}
\end{figure}

\begin{figure}
\includegraphics[width=\columnwidth]{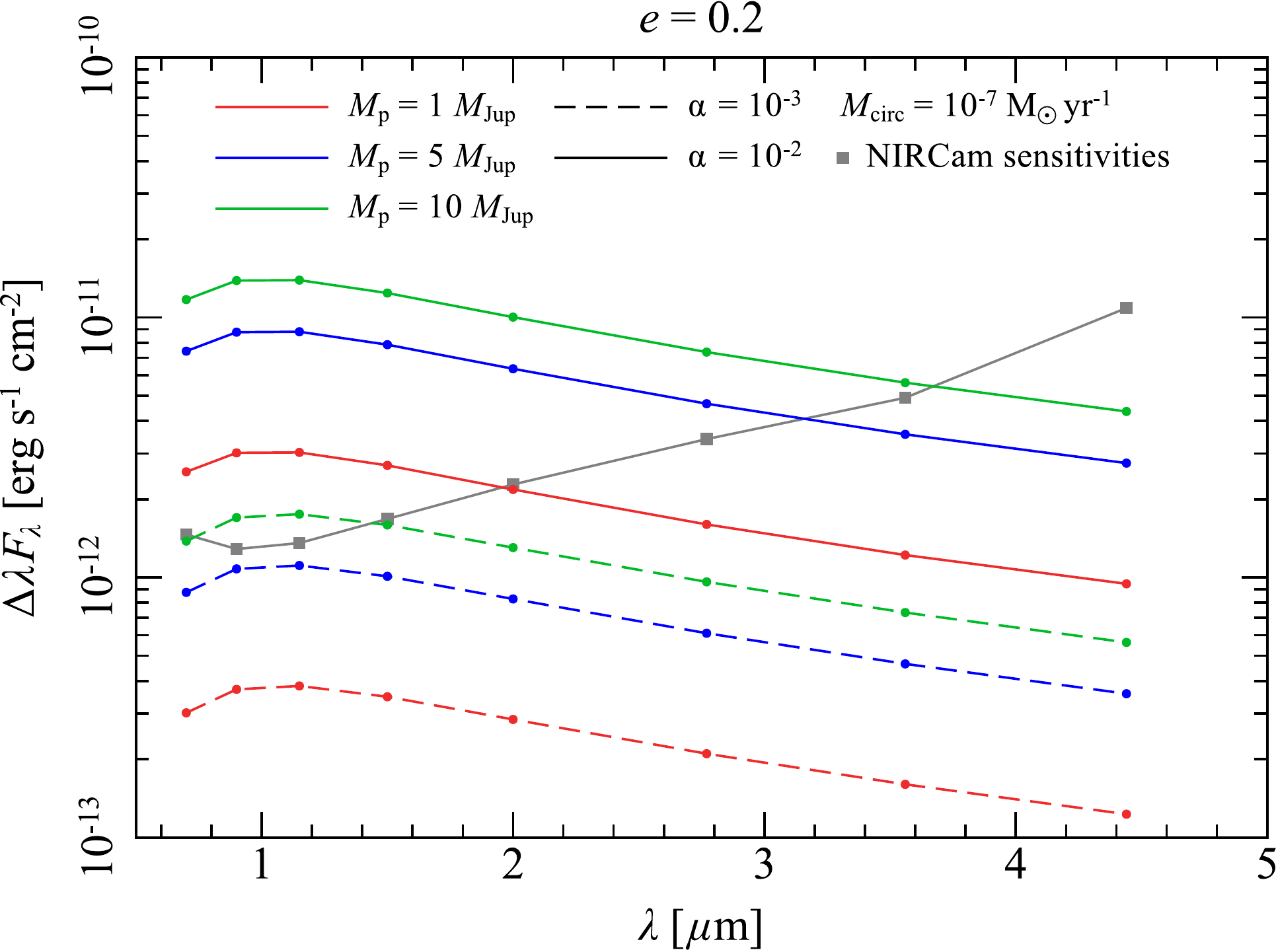}
\caption{As Figure \ref{fig:3} but for planets with eccentricity $e = 0.2$.}
\label{fig:4}
\end{figure}

I also neglect realistic viscosity parameterisations and consideration of vertical disc layers \citep[e.g.][]{lubowmartin12,lubowmartin13} as would be appropriate for a CPD with a dead zone. These are much less viscous than $\alpha$ discs and as Figures \ref{fig:3} and \ref{fig:4} show, low viscosity drastically reduces the observability of the eccentric modulation so a dead zone model would produce less optimistic results.

Other aspects I do not investigate include the injection radius $R_{\mathrm{inj}}$ and the strength and radius of the tidal torque $dT_{\mathrm{gr}}/dM$. The effect of changing both was well tested by \citet{martinlubow11}, and while they do alter the structure of the CPD this should be robust against eccentricity as parameterised in the 1D model. As I do not resolve below $0.4 R_{\mathrm{Hill}}$ in the SPH simulations used to calibrate the accretion rates, it is not clear from these what the appropriate value for $R_{\mathrm{inj}}$ is in a real disc. This depends on where the infalling gas shocks and how efficiently it cools, analogous to the same mechanism in binary accretion \citep[e.g.][]{clarke12}. Resolving this issue will require 3D radiation hydrodynamic simulations.

The flux calculations here are also greatly simplified, especially when compared to recent work by \citet{zhu14} who calculated full SED models for discs around planets on circular orbits. In this work I focus on the level of variability produced by an eccentric planet, rather than a faithful prediction of the full SED. This variability should to first order be independent of the details of the SED, so I choose not to use a self-consistent temperature calculation in the model, instead using a fixed $H/R$.

While it is not clear what the $H/R$ should be for a CPD \citet{martinlubow11} show that for high mass planets lower values are more likely, but $H/R \sim 3$ is a common outcome of simulations \citep[e.g.][]{ayliffebate09b} and so I adopt it here. Altering $H/R$ has the same effect as reducing the viscosity (remembering that $\nu = \alpha H^2\Omega$), and while for a low enough $H/R$ this alters the shape of the light curve in Figure \ref{fig:2} it does not affect $\Delta\lambda F_{\lambda}$. Further, the flux emission shown in Figure \ref{fig:2} is roughly consistent with the SED models of \citet{zhu14} at 10 $\mu$m when corrected for distance (for $\alpha = 10^{-3}$ and $\dot{M}_{\mathrm{circ}} = 10^{-7}$ at $d = 100$ pc, log $\lambda F_{\lambda} = -10.7$; compare with Figure 1, bottom left panel, from that paper, with $M_{\mathrm{p}}\dot{M} = 10^{-4}M_{\mathrm{Jup}}^2\,$ yr$^{-1}$, for which log $\lambda F_{\lambda} \simeq -10.5$ at $1.5\,\mu$m).

I also neglect to include the flux contribution from magnetospheric accretion directly onto the planet in my calculations. For cases where the CPD viscosity is high ($\alpha = 0.01$), the viscous timescale in the at the injection radius \citep[$t_{\nu} = R_{\mathrm{inj}}^2 / 12 \nu$;][]{pringle81} is shorter than the orbital period of the planet so the accretion of mass onto the planet is strongly periodic. For lower values of $\alpha$ the periodicity in the mass accretion is at an extremely low level. However, \citet{zhu14} showed that for this to have a significant effect on the SED of the object, the planet's magnetic field strength needs to be of the order $100 - 1000$ G -- compared with the value of 4.28 G for Jupiter, this is unlikely to be the case, even considering that younger planets likely have stronger magnetic fields.

\subsection{Interpretation}\label{discussion:interpretation}

The primary result that is an eccentric planet modulates the flux emission from its CPD. To be observable, high eccentricities, accretion rates and CPD viscosites are required. While the $\alpha$-viscosity used here is likely inaccurate for describing a CPD, the correct treatment is not obvious. There is a growing body of work investigating the effect of different levels of CPD viscosities \citep[e.g.][]{lubowmartin12,lubowmartin13,buetal13,gresseletal13,szulagyietal14,keithwardle14} no consensus has yet emerged.

The primary uncertainty here is the CPD temperature. How the gas in the planet's vicinity is heated obviously has strong implications, as this controls both the viability of the MRI and to what extent self-gravity can play a role in driving turbulent viscosity in the disc. If the planet is shielded from the central star, perhaps by an optically thick inner gap edge, or simply by a large amount of co-orbital gas, then the CPD will be cold and relatively inviscid \citep{lubowmartin13,szulagyietal14}, unless it is so cold that it becomes unstable to self-gravity at which point gravitoturbulence can drive the viscosity up again \citep{keithwardle14}.

If the disc is hot (for example if its orbit is close to the star) then it may be MRI active throughout and capable of driving rapid accretion through a high turbulent viscosity \citep[e.g.][]{gresseletal13}. Hydrodynamical simulations including radiative physics have shown that CPDs are expected to have large aspect ratios, with $H/R \sim 0.3-0.6$ \citep{ayliffebate09b}, and while this may indicate that they should be hot and possibly vulnerable to MRI, these simulations are very sensitive to assumptions about grain opacity which may not be accurate for planet-forming discs.

It seems that we require direct and unambiguous observations to break this degeneracy. SED models by \citet{zhu14} show the multi-band IR observations of a CPD can help constrain its properties and begin to probe the rest of the circumstellar disc structure.

Observing this eccentric modulation would have strong implications for our protoplanetary disc conditions. While resonant disc interactions can grow eccentricity \citep[e.g.][]{goldreichtremaine80,papaloizouetal01}, The range of disc parameters that permit this is uncertain \citep{ogilvielubow03,goldreichsari03,massetogilvie04}. In cases where the eccentricity does not grow in this way, it is efficiently damped \citep{dunhilletal13,tsangetal14}, so we can ascribe the eccentricity of an embedded planet to these torques with a high level of confidence.

Interestingly, \citet{tsangetal14} have shown that a planet gap heated by its parent star is required for this mechanism to operate -- the same situation required for the MRI to drive a high viscosity in the CPD. This increases the likelihood of observational confirmation that discs can grow planetary eccentricity.

An obvious complication in actually observing this variability is the fact that young stars with discs are known to be inherently variable in the near-Infrared, independent of any emission from circumplanetary discs \citep[e.g.][]{moralescalderonetal09,moralescalderonetal11}, often with quasi-periodicities on the order of tens of days \citep{rebulletal14,staufferetal14}. Observations with \textit{Spitzer} show that this variability typically has a magnitude of $\Delta\lambda F_{\lambda} \sim 5\times 10^{-12}$ erg s$^{-1}$ cm$^{-2}$ at wavelengths $\lambda = 3.6 - 4.5$ $\mu$m \citep[e.g.][]{rebulletal14}. This is at a level comparable with the most optimistic variability due to an eccentric planet in my models. However typical variability in these objects, be it due to stellar pulsations, accretion events or variably obscuration, is rarely of a purely periodic nature \citep{codyetal14} while the signal from an eccentric CPD should be.

With future \textit{ALMA} observations it may be possible to discern the exact period of an accreting giant planet by resolving the gap it creates in the disc (see e.g. the recent image of HL Tau). Thus, knowing the period of the planet \textit{a priori} will make monitoring for a periodic signal from its CPD much simpler. Finding such a signal will still be difficult though, especially given the unknown real performance of \textit{JWST} when it launches. Indeed, it is likely that monitoring on long timescales at the required sensitivities will not be possible due to eventual decay in instrument performance. It is fortunate, then, that the most likely candidates for eccentricity growth from disc torques are those orbiting at small radii ($R \lesssim 1$ au) where the disc is directly heated by the star \citep{tsangetal14}.

\section{Conclusions}\label{sec:conclusions}

I have used simple 1D models of circumplanetary discs around eccentric planets to calculate the orbital modulation of emission from the disc using classical accretion disc assumptions. For a disc around a planet forming in the nearby TW Hya Association, the level of modulation is above the minimum specification sensitivity required for \textit{JWST}'s NIRCam instrument, for certain disc parameters and orbital eccentricities. If these minimum sensitivities are accurate, then an accretion rate onto the circumplanetary disc of $\dot{M} \sim 10^{-7}\,\mathrm{M}_{\sun}$ yr$^{-1}$ is required for any variability to be observed.

For all the planet masses studied here, $1 \lesssim M_{\mathrm{p}} \lesssim 10\,M_{\mathrm{Jup}}$, a high viscosity $\alpha \sim 10^{-2}$ is required for the modulation to be above the minimum observable limit, except for the most massive planets which are just above the NIRCam sensitivity limits at $0.9$ and $1.15$ $\mu$m at lower viscosity for $e = 0.2$ (see Figure \ref{fig:4}).

I conclude that while these parameters (especially the viscosity) are at the high end of what is realistic, they are still within the bounds of current observational and theoretical limits. Further modelling, in the form of full 3D hydrodynamic simulations and more sophisticated SED work, is required to form an accurate observability study of this effect.

\section*{Acknowledgments}

I would like to thank Richard Alexander, Jorge Cuadra and Rebecca Martin for very helpful comments on early versions of the manuscript. I also thank the anonymous referee for their thoughtful comments. I acknowledge support from from ALMA CONICYT grant 311200007 and CONICYT PFB0609.

\bibliography{mnrasmnemonic,references}
\bibliographystyle{mnras}

\label{lastpage}

\end{document}